\documentclass[runningheads]{llncs}
\usepackage{graphicx}
\usepackage{subfigure}
\usepackage{booktabs}
\begin{document}
\title{OctopusNet: A Deep Learning Segmentation Network for Multi-modal Medical Images}
\author{Yu Chen\inst{1, 2}\thanks{This work was done when Yu Chen was an intern at YouTu Lab} \and
Jiawei Chen\inst{2} \and
Dong Wei\inst{2} \and Yuexiang Li\inst{2} \and Yefeng Zheng\inst{2}}
% Yu Chen \and Yuexiang Li \and Jiawei Chen \and Yefeng Zheng
\authorrunning{Y. Chen et al.}
% Y. Chen et al.
\institute{Nanjing University, Nanjing, China\\
\and YouTu Lab, Tencent, Shenzhen, China\\
\email{vicyxli@tencent.com}
}
\maketitle              % typeset the header of the contribution
\begin{abstract}
	Deep learning models, such as the fully convolutional network (FCN), have been widely used in 3D biomedical segmentation and achieved state-of-the-art performance. Multiple modalities are often used for disease diagnosis and quantification. Two approaches are widely used in the literature to fuse multiple modalities in the segmentation networks: early-fusion (which stacks multiple modalities as different input channels) and late-fusion (which fuses the segmentation results from different modalities at the very end). These fusion methods easily suffer from the cross-modal interference caused by the input modalities which have wide variations. To address the problem, we propose a novel deep learning architecture, namely OctopusNet, to better leverage and fuse the information contained in multi-modalities. The proposed framework employs a separate encoder for each modality for feature extraction and exploits a hyper-fusion decoder to fuse the extracted features while avoiding feature explosion. We evaluate the proposed OctopusNet on two publicly available datasets, i.e. ISLES-2018 and MRBrainS-2013. The experimental results show that our framework outperforms the commonly-used feature fusion approaches and yields the state-of-the-art segmentation accuracy.

	\keywords{Medical image segmentation  \and Deep learning \and Multi-modal images.}
\end{abstract}
\section{Introduction}
Recent years have witnessed the rapid development of deep learning technique. Deep learning models have been widely used for medical image segmentation and achieved impressive performance \cite{Pereira2018AdaptiveFR,Shenh_no1,no_19}. Compared with natural images, medical images, e.g. computed tomography (CT) and magnetic resonance imaging (MRI), often have a lot of scanning protocols in its toolbox and each protocol may reveal a different property (often complementary to other protocols) of the underlying tissue. For examples, to assess ischemic stroke lesion, three modalities using perfusion imaging are commonly captured, i.e. cerebral blood volume (CBV), cerebral blood flow (CBF), and time to peak of the residue function (Tmax). Those modal images may contain different clinical interpretation.

\begin{figure}[!tb]
	\begin{minipage}[t]{0.49\linewidth}
		\centering
		\includegraphics[width=2.0in]{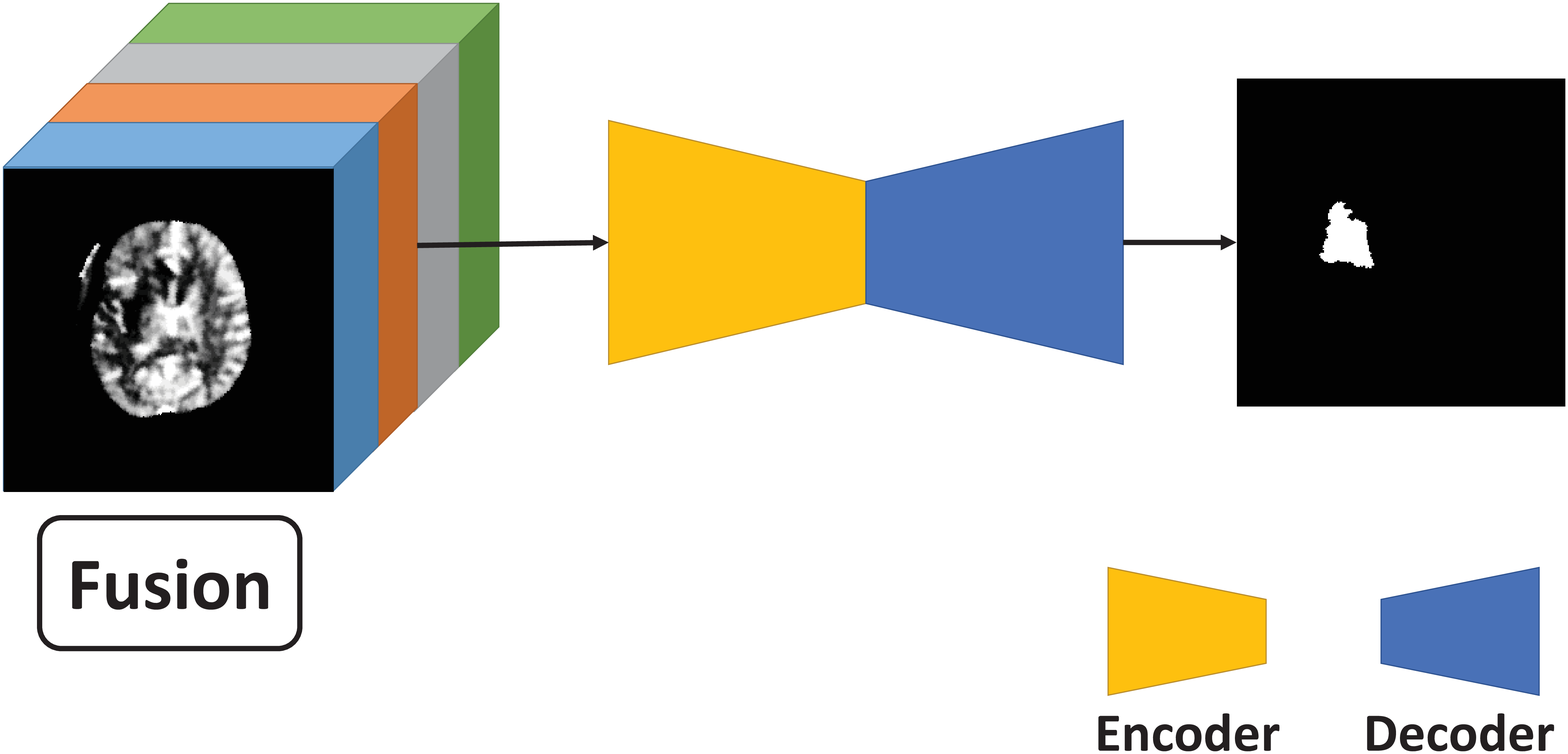}
		\scriptsize{\\(a) Early-fusion: fusion in the input stage\\}
	\end{minipage}
	\begin{minipage}[t]{0.49\linewidth}
		\centering
		\includegraphics[width=2.0in]{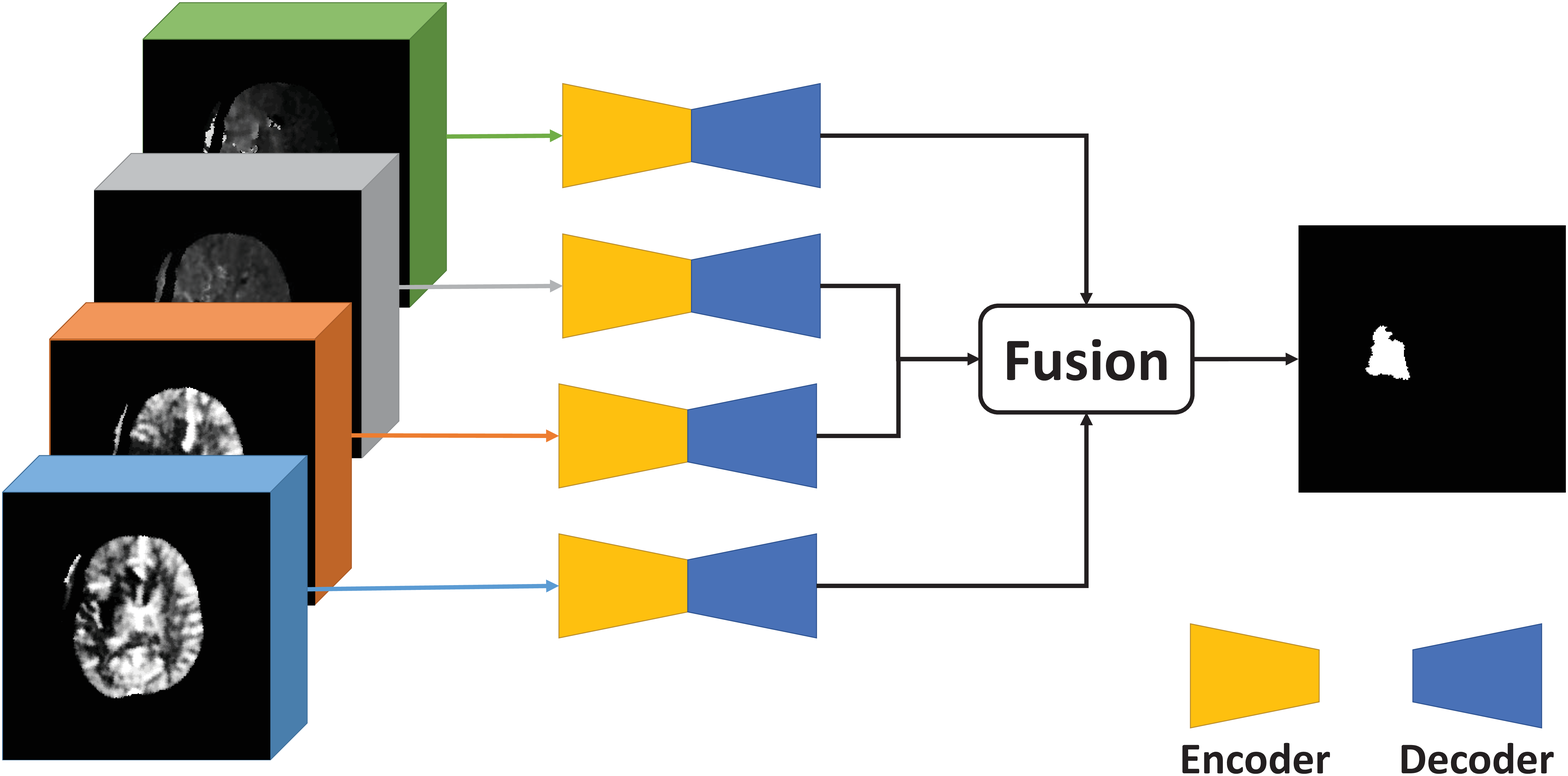}
		\scriptsize{\\(b) Late-fusion: fusion in the output stage\\}
	\end{minipage}
	\begin{minipage}[t]{\textwidth}
		\centering
		\includegraphics[width=0.95\textwidth]{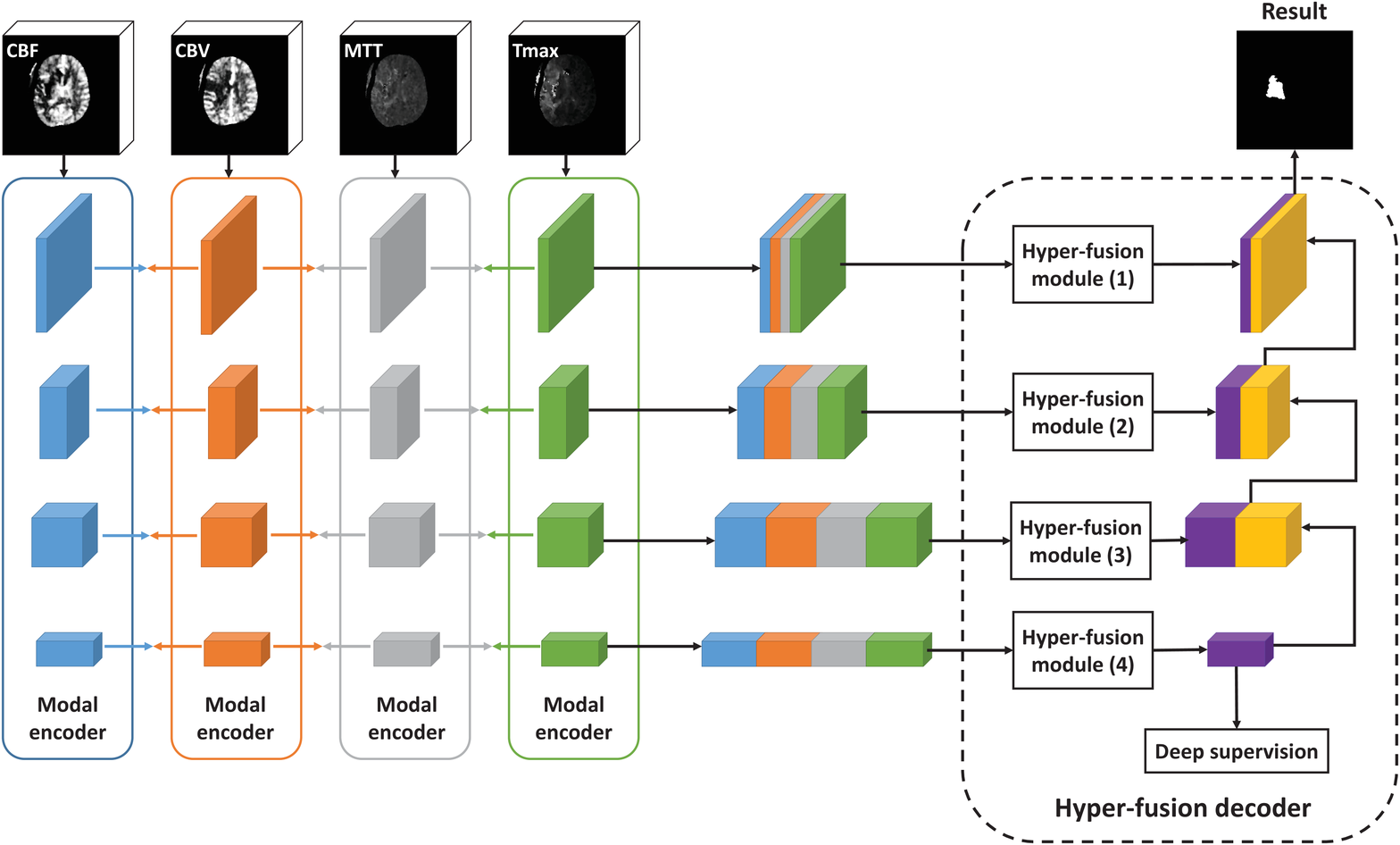}
		\scriptsize{\\(c) OctopusNet}
	\end{minipage}
	\caption{Different fusion approaches, including early-fusion (a), late-fusion (b) and the architecture of our OctopusNet (c).} \label{fig1}
\end{figure}

To exploit the multi-modal medical data, two fusion approaches are widely-used by current deep learning networks, i.e. stacking multiple modalities as different input channels (early-fusion, Fig.~\ref{fig1}(a)) \cite{Pereira2018AdaptiveFR,Shenh_no1,Wangl_no1} and fusing the outputs of networks from different modalities (late-fusion, Fig.~\ref{fig1}(b)) \cite{no_19,Wuz_no1}. Neither fusion approach is optimal in using the complementary information from multiple input modalities. Take the perfusion CT for ischemic stroke lesion segmentation as an example. As shown in Fig.~\ref{fig1}(c), four modalities, i.e. CBV, CBF and MTT, and Tmax, are captured. It can be observed from the four modalities that the lesion area in CBV and CBF is darker compared to the normal area, while it is lighter in the modalities of MTT and Tmax. Consequently, the information in different modalities may be wrongly fused if we simply adopt the early-fusion approach. On the other hand, although the late fusion approach adopts the separate encoder-decoder for each modality, the whole network is computational expensive and difficult to converge.

In this paper, we propose a novel segmentation network, namely OctopusNet, which effectively leverages the information contained in multi-modal medical images. Instead of fusing multi-modal images at the input stage, we exploit an individual encoder for each modality and fuse the feature maps generated by middle stages of the network, which specifically extracts features from each modality and explicitly considers the correlations between different modalities. As the modalities are separately encoded, the proposed OctopusNet adopts a novel feature fusion module, namely a hyper-fusion decoder, to merge the feature maps and avoid feature explosion. Extensive comparison experiments are conducted on multi-modal datasets. The results demonstrate the outstanding segmentation performance of the proposed OctopusNet.

\section{OctopusNet}
In this section, we introduce the detailed information of our OctopusNet. The framework of our OctopusNet is shown in Fig.~\ref{fig1}(c). The colored cubes refer to the feature maps generated at different stages of the framework. Our OctopusNet addresses the problem of cross-modal interference, occured in early-fusion, by extracting features from the modalities using separate modal encoders, which can be any CNN architecture, e.g. VGG \cite{no_10}, ResNet \cite{no_7} or DenseNet \cite{no_11}. As shown in Fig.~\ref{fig1}(c), the feature maps generated at different stages of modal encoders are concatenated and fed to the hyper-fusion decoder. The proposed decoder uses hyper-fusion modules to fuse the feature maps from different modalities and avoid the problem of feature explosion. Compared to the late-fusion approach with four separate decoders, the hyper-fusion decoder more effectively fuses cross-modal information and reduces the computational cost. The decoder upsamples the high-level low-resolution feature maps back to the original resolution in the same way as \cite{no_3}, and yields the segmentation result.

\subsection{Modal encoder}
As aforementioned, the modal encoder can be chosen from any widely-used network architectures, e.g. DenseNet. In our experiments, DenseNet-161 usually yields better segmentation accuracy compared to that of VGG and ResNet. Therefore, we take DenseNet-161 as an example to illustrate the pipeline of extracting feature maps from different modalities. The detailed information of network architecture of DenseNet-161 can be found in \cite{no_11}. The colored cubes in each modal encoder in Fig.~\ref{fig1}(c) are the feature maps generated by different stages of DenseNet-161, which correspond to the ones from Dense Block (1) - (4). To better leverage the explicit information contained in different modalities, all the extracted feature maps are concatenated together and fed to the hyper-fusion decoder for feature fusion and distilling.

\subsection{Hyper-fusion decoder}
We propose a novel hyper-fusion decoder to decode and upsample the high-level low-resolution feature maps back to the original resolution of input and yield the segmentation result. As shown in Fig.~\ref{fig1}(c), the decoder adopts a {\itshape hyper-fusion module} for feature distilling, {\itshape deep supervision} for better training convergence, and concatenates the fused feature maps (purple cubes) to the upsampled ones (yellow cubes) to produce segmentation result.

\vspace{0.1in}
\noindent{\bf Hyper-fusion module.}
As the network goes deeper, the number of feature maps increases. Consequently, the concatenation of multi-modal feature maps easily causes a problem of feature explosion. For example, assuming there are N modalities as input, the number of concatenated feature maps generated by Dense Block (4) of modal encoders is $2208 \times N$, which requires high cost of computation and memory consumption. The hyper-fusion module (1) - (4) in Fig.~\ref{fig1}(c) is a $1 \times 1$ convolution, which has the same number of channels to that of feature maps from Dense Block (1) - (4). Therefore, the N concatenated feature maps can be accordingly fused and compacted to single ones using hyper-fusion modules (the purple cubes in Fig.~\ref{fig1}(c)).

\vspace{0.1in}
\noindent{\bf Deep supervision.}
The proposed OctopusNet is an end-to-end framework, which means the multiple modal encoders and hyper-fusion decoder are simultaneously trained and updated. However, the networks adopted for modal encoders are usually extremely deep, resulting in a difficulty for training convergence only using a single supervision signal at the very end of a long pipeline. Hence, we added a weak supervision signal to the deepest node of OctopusNet, i.e. the elongated purple cube at the bottom. Assuming DenseNet-161 is adopted as the modal encoder, the size of bottom purple cube is $2208 \times 7 \times 7$. In this situation, a $1 \times 1$ convolution is used to transform the cube to $1 \times 7 \times 7$ and the original supervision signal is resized from $224 \times 224$ to $7 \times 7$ to be as weak supervision.

\section{Experiments}
We evaluate the performance of the proposed OctopusNet on publicly available datasets from two challenges, namely ISLES-2018\footnote{http://www.isles-challenge.org/} and MRBrainS-2013\footnote{http://mrbrains13.isi.uu.nl/index.php}, and compare with the early-fusion and late-fusion approaches. Though our result is competitive to the top-performance of the ISLES-2018 challenge, the purpose of the experiments is not to win the challenges. Our main purpose is to demonstrate the effectiveness of the proposed fusion approach and compare with other widely used alternative strategies. Our approach is complementary to, and can be easily integrated into, other FCN based multi-modal segmentation approaches.

\subsection{Datasets}
\vspace{0.1in}
\noindent{\bf ISLES-2018.}
Ischemic stroke lesions segmentation (ISLES) is a competition consecutively held since 2015 \cite{no_14}. In ISLES-2018, the challenge organizer released a new dataset, which is composed of six modalities, including diffusion weighted imaging (DWI) MRI, computed tomography (CT) and four perfusion scans, i.e. mean transit time (MTT), time to peak of the residue function (Tmax), cerebral blood flow (CBF) and cerebral blood volume (CBV). The ISLES-2018 competition provides 94 sets of multi-modal data for training and 62 sets for test. As the ground truth of the test set is not available, participants need to submit their prediction to the online system for performance evaluation.

The dataset has two main challenging issues. First, the appearances of lesion areas among different modalities are widely varied. The lesion area in MTT and Tmax is brighter than the normal area, while it is dark in the modalities of CBF and CBV, as shown in Fig.~\ref{fig1}(c). Second, the ISLES-2018 data does not have a uniform size. Though each slice has a fixed size of $256 \times 256$ pixels, the number of slices contained in a volume varies from 2 to 22. Most ISLES-2018 volumes only have two slices, which presents a difficulty to adapt a 3D segmentation framework to the dataset.

\vspace{0.1in}
\noindent{\bf MRBrainS-2013.}
The MRBrainS-2013 dataset contains five sets of multi-modal brain images, in which the brain tissues, i.e. gray matter, white matter and cerebrospinal fluid, are fully annotated. Three registered modalities, i.e. T1-weighted scan (T1), T1-weighted inversion recovery scan (T1\_IR) and T2-weighted fluid attenuated inversion recovery scan (T2\_FLAIR) are provided. The volumes of the dataset are in an uniform size of $240 \times 240 \times 48$ voxels.

\vspace{0.1in}
\noindent{\bf Implementation details.}
The proposed OctopusNet may have different architectures regarding to the input data. As most ISLES-2018 data has a couple of slices, we develop a 2.5D OctopusNet instead of 3D. Three consecutive slices from a volume are extracted and fed to 2D modal encoders as inputs. The first and last slices of the volume are duplicated for padding. In this setting, modal encoders can be pretrained on the ImageNet dataset for better training convergence. Our OctopusNet is implemented using PyTorch. The initial learning rate is set to 0.7 and divided by 10 after every 35 epochs. The network is optimized by stochastic gradient descent (SGD). The used datasets have different number of modalities. Consequently, the proposed OctopusNet involves different numbers of modal encoders for the ISLES-2018 and MRBrainS-2013.

Take the ISLES-2018 as an example. As the ISLES-2018 test set does not provide the DWI modality, five original modalities, i.e. CT, MTT, Tmax, CBF and CBV, are adopted as input for OctopusNet. Furthermore, the lesion area is not clearly visible in the modalities of CT, CBF and CBV. Hence, these three modalities are enhanced by histogram equalization. Finally, five original modalities and three enhanced modalities are fed to the OctopusNet\footnote{This network has an octopus shape with a body (the decoder) and eight arms (the encoders). This is where the name, OctopusNet, comes from.}. The input size of each modal encoder is $256 \times 256 \times 3$.
% We notice that there are some similar network architectures proposed for other challenges of MICCAI 2018. It is worthwhile to mention that we originally develop the OctopusNet and conduct comprehensive experiments to evaluate its performances.
%\cite{Pereira2018AdaptiveFR,Shenh_no1,Wangl_no1} \cite{no_19,Wuz_no1}
\subsection{Performance Analysis}
We perform a five-fold cross validation on the ISLES-2018 and MRBrainS-2013 training set to evaluate the performance of our OctopusNet. All the experiments are repeated three times to reduce the influence caused by random nature of network training. Hence, the results reported in the paper are the the average results of three repeated experiments. For the convenience of comparison, the frameworks using baseline fusion approaches (early- and late-fusion) in our experiments are in the same setting to that of OctopusNet, e.g. the input format of different modalities. Henceforth, the fusion approach adopted in our OctopusNet is named as Octopus-fusion. The Dice coefficient, which measures the spatial overlap index between the segmentation results and ground truths, is adopted as the metric to evlauate the segmentation accuracy.

\begin{table}[!tb]
	\caption{Dice coefficient (\%) of lesion areas of ISLES-2018 (average of five-fold cross validation).}\label{tab2}
	\centering
	\begin{tabular}{lccc}
		\toprule
		{ }                                           & VGG-16 \cite{no_10} & ResNet-50 \cite{no_7} & DenseNet-161 \cite{no_11} \\
		\toprule
		{\bfseries Single modality (Tmax)}            & 44.97               & 44.03                 & 45.83                     \\
		\hline
		{\bfseries Early-fusion}                      & 53.38               & 53.99                 & 53.82                     \\
		\hline
		{\bfseries Late-fusion}                       & 53.73               & 55.39                 & 53.86                     \\
		\hline
		{\bfseries Octopus-fusion}                    & {\bfseries 55.71}   & {\bfseries 57.33}     & {\bfseries 57.72}         \\
		\hline
		{\bfseries Octopus-fusion + deep supervision} & -                   & -                     & 57.90                     \\
		\bottomrule
	\end{tabular}
\end{table}

\vspace{0.1in}
\noindent{\bf Results on ISLES-2018.}
As aforementioned, the modal encoder can be chosen from widely used deep learning networks. To evaluate the generalization capability of Octopus-fusion, several network architectures, e.g. VGG-16 \cite{no_10}, ResNet-50 \cite{no_7} and Dense-Net-161 \cite{no_11}, are adopted as the modal encoder and trained with different fusion approaches on the ISLES-2018 dataset. The results are listed in Table~\ref{tab2}. To evaluate the improvement produced by the usage of multi-modal images, we also report segmentation accuracy using a single modality. Due to the space limit, Table~\ref{tab2} only lists the result of the best single modality (i.e., Tmax). Due to the lack of information contained in extra modalities, the frameworks using single modality only yield Dice coefficients around 44\%, which are about 9\% lower than that of multi-modal frameworks. For the early-fusion approach, Table~\ref{tab2} shows that the accuracies of all three backbone networks are quite similar with the deep networks (i.e., DenseNet-161 and ResNet-50) slightly outperforming the shallow network of VGG-16.

For the late-fusion approach, as it involves multiple encoder-decoder architectures for different modalities, the explicit information contained in multi-modal data can be better extracted. Hence, accuracy of the late-fusion approach surpasses that of early-fusion with the same modal encoders. However, DenseNet-161 only gains marginal improvement, i.e. 0.04\%, by switching from early-fusion to late-fusion. The reason for that is the network depth of DenseNet-161 is extremely deep, which makes it difficult to simultaneously well train multiple fully convolutional DenseNet-161 branches in the late-fusion approach. Compared to the results of early- and late-fusion, our Octopus-fusion approach significantly boosts the accuracy of modal encoders. The best lesion segmentation result is achieved by the Octopus-fusion DenseNet-161, i.e. a Dice of 57.72\%. By adding the {\itshape deep supervision} signal, the segmentation accuracy is further increased to 57.90\%, which is 2.51\% higher than that of the best-performance among benchmarking algorithms (late-fusion with ResNet-50).

{\bf ISLES-2018 Challenge.}
We participated the ISLES-2018 competition. The proposed OctopusNet using DenseNet-161 achieved an average Dice of 48\%, which ranked the third-place of ISLES-2018 challenge\footnote{https://www.smir.ch/ISLES/Start2018}. We notice that, for all participating teams, there is a gap between validation and test accuracy. One possible reason is that the test set contains more small lesions, where are difficult to segment accurately for all algorithms. Additionally, the top approaches reported that they used extra modalities, e.g. 4D perfusion CT (ranked 1$^{st}$ with Dice of 51\%) and synthesized DWI (ranked 2$^{nd}$ with Dice of 49\%), which were not adopted by our OctopusNet.

\vspace{0.1in}
\noindent{\bf Results on MRBrainS-2013.}
We also conduct experiments on MRBrainS-2013 to compare the performances of different fusion approaches for the task of brain tissue segmentation. The three original modalities of MRBrainS-2013 are directly employed as input for the proposed OctopusNet. The best-performer on ISLES-2018, i.e. DenseNet-161, is adopted as the backbone of modal encoder. The input size of each modal encoder is $240 \times 240 \times 3$. The Dice coefficients for different tissues, including CSF, gray matter and white matter, produced by different fusion approaches are listed in Table~\ref{tab3}. The average Dice (Ave. Dice) is calculated by averaging the Dice coefficients of three tissues.

\begin{table}[!tb]
	\caption{Dice coefficients (\%) yielded by different fusion approaches for each brain tissue of MRBrainS-2013 (average of five-fold cross validation).}\label{tab3}
	\centering
	\begin{tabular}{lcccc}
		\toprule
		{ }                                     & CSF               & Gray matter       & White matter      & Ave. Dice         \\
		\toprule
		{\bfseries Single modality (T1)}        & 78.46             & 81.37             & 85.69             & 81.84             \\
		\hline
		{\bfseries Single modality (T1\_IR)}    & 75.95             & 77.51             & 81.92             & 78.46             \\
		\hline
		{\bfseries Single modality (T2\_FLAIR)} & 74.00             & 75.63             & 77.32             & 75.65             \\
		\hline
		{\bfseries Early-fusion}                & 79.05             & 80.58             & 83.56             & 81.07             \\
		\hline
		{\bfseries Late-fusion}                 & 79.16             & 81.41             & 84.71             & 81.76             \\
		\hline
		{\bfseries Octopus-fusion}              & {\bfseries 80.59} & {\bfseries 82.12} & {\bfseries 86.05} & {\bfseries 82.92} \\
		\bottomrule
	\end{tabular}
\end{table}

The framework using single modality is also evaluated for comparison. It is interesting to see that, for the gray matter and white matter, the best single modality (T1) produces even higher segmentation accuracy than early-fusion. The reason for that may be the physicians mark the annotation of the gray and white matter primarily using the T1 scans, while the T1\_IR and T2\_FLAIR scans usually provide additional information for the annotation of outer border of CSF and white matter lesion, respectively. Therefore, most information contained in the extra modalities, i.e. T1\_IR and T2\_FLAIR, may be seen as noises for the brain tissue segmentation. The late-fusion approach yields similar average segmentation accuracy to using T1 only (81.76\% vs. 81.84\%). The reason for that may be the post-fusion approach performs information fusion too late; Therefore, it can not fully utilize the complementary information among multiple modalities. Oppositely, by using the proposed Octopus-fusion, the average segmentation accuracy increases to 82.92\% and improvement is observed for all tissues, which illustrates that our Octopus-fusion can effectively extract useful information from each modality and prevent the cross-modal interference caused by irrelevant information. An additional observation is that CSF consistently benefits from multi-modality fusion using any fusion strategy, which is concordant to the annotation process of physicians. Again, Octopus-fusion achieves the largest boost in segmentation accuracy of CSF, i.e., $+2.13\%$.

\section{Conclusion}
In this paper, we presented a novel deep learning network architecture, namely OctopusNet, for multi-modal medical image segmentation. The proposed OctopusNet adopted a separate modal encoder for each modality to explicitly extract features and a hyper-fusion decoder to fuse the features, avoiding the problem of feature explosion. The proposed OctopusNet was evaluated on two publicly available datasets. The experimental results demonstrated that our OctopusNet was a general network architecture, which can provide excellent performance for various segmentation tasks of multi-modal medical data.


\end{document}